\documentclass[12pt]{spieman}  
\usepackage{amsmath,amsfonts,amssymb}
\usepackage{graphicx}
\usepackage{setspace}
\usepackage{tocloft}
\usepackage{lineno}
\usepackage{authblk} 
\usepackage{url,hyperref,lineno,microtype,subcaption}
\usepackage{tabularx}
\usepackage{amsmath}
\usepackage{xcolor}
\usepackage[numbers]{natbib}
\usepackage{wrapfig}
\usepackage{underscore}
\usepackage{amssymb} 
\usepackage{tikz,lipsum,lmodern}
\usepackage[most]{tcolorbox}
\usepackage{placeins}

\title{A Science Traceability Matrix for micro-arcsecond resolution X-ray imaging}

\author[a,b,c*]{Jenna M. Cann}
\author[a]{Kimberly A. Weaver}
\author[d]{Ryan W. Pfeifle}
\author[a,e]{Jeffrey D. McKaig}
\author[a,b,c]{Teresa Monsue}
\author[a,c,g]{Laura D. Vega}
\author[a,c,g]{Ronald Gamble}
\author[i]{Miranda McCarthy}
\author[a,c,h]{Kyla Mullaney}
\author[a,e]{Kelly E. Whalen}
\author[a,c,g]{Erini Lambrides}
\author[j]{Atul Mohan}
\author[k]{Isabella Carlton}
\author[i]{Emma Kleiner}
\author[l]{Kaylee DeGennaro}
\author[a,c,g]{Mainak Singha}
\author[a,c,h]{Andrea Astorga Bedoya}
\author[a,c,h]{Neshini Sathishkumar}

\affil[a]{NASA Goddard Space Flight Center, Greenbelt, MD 20771, USA}
\affil[b]{Center for Space Science and Technology, University of Maryland, Baltimore County, 1000 Hilltop Circle, Baltimore, MD 21250, USA}
\affil[c]{Center for Research and Exploration in Space Science and Technology (CRESST II), NASA/GSFC, Greenbelt, MD 20771, USA}
\affil[d]{U.S. Naval Observatory, 3450 Massachusetts Avenue NW, Washington, DC 20392, USA}
\affil[e]{Oak Ridge Associated Universities (ORAU), 100 Orau Way, Oak Ridge, TN 37830,
USA}
\affil[f]{Catholic University of America, 620 Michigan Ave. NE, Washington, DC 20064,
USA}
\affil[g]{University of Maryland, College Park, 3972 Campus Dr. College Park, MD 20742,
USA}
\affil[h]{Southeastern Universities Research Association (SURA), 1401 I St NW, Suite 500, Washington, DC 20005}
\affil[i]{City University of New York, 205 East 42 St, New York, New York 10017, USA}
\affil[j]{National Astronomical Observatory of Japan, 2-21-1 Osawa, Mitaka, Tokyo 181-8588, Japan}
\affil[k]{University of Utah, 201 Presidents Circle, Salt Lake City, UT 84112, USA}
\affil[l]{Brown University, 182 Hope Street, Providence, RI 02912}

\cftpagenumbersoff{figure}
\cftpagenumbersoff{table} 
\begin{document} 
\maketitle

\begin{abstract}
Our ability to understand the highest energy processes and environments in astronomy is directly tied to our ability to probe their physical processes at the smallest scales. Despite transformative progress in high-angular resolution instruments across the radio through optical, angular resolution within the X-rays has plateaued around $\sim0.1-1''$, and therefore our window into the innermost regions of the high energy Universe lies primarily within spectral, timing, and/or polarimetric studies rather than direct imaging. This represents a significant barrier to gaining a robust understanding of key science questions, including the mechanisms of accretion and jet launching, the accretion activity and angular momentum losses of binary active galactic nuclei (AGN) leading to gravitational wave events, and the space weather events that may influence atmospheric evolution and prebiotic chemistry on potentially habitable exoplanets. The Accretion Explorer Interferometer (AEI) is a mission concept for X-ray imaging that could achieve up to $\mu$as resolution, $\sim5$--$6$ orders of magnitude better than current instrumentation. Here we present the scientific discovery space enabled by such an advance, and highlight areas of uncertainty and open questions that are critical to investigate for future development of this concept.

\end{abstract}


{\noindent \footnotesize\textbf{*}Jenna Cann,  \linkable{jcann@umbc.edu}, \linkable{jenna.cann@nasa.gov} }

\begin{spacing}{2}   

\section{Introduction}
\label{sec:intro}  
Despite significant technological advancement across multi-wavelength observatories, our only window into the X-ray universe lies within unresolved images. In fact, Chandra, the highest resolution X-ray imager currently flying, can attain angular resolutions of only $\sim0.5-1''$, lagging significant behind its multi-wavelength counterparts that regularly achieve angular resolutions in the milli-arcsecond (mas; HST, JWST, VLA, ALMA) and even micro-arcsecond ($\mu$as; EHT) range. Even next generation X-ray mission concepts such as the Advanced X-ray Imaging Satellite \citep[AXIS;][]{axis} and the Lynx X-ray Observatory \citep[Lynx;][]{lynx} propose limited advancements in angular resolution, either maintaining the status quo at $\sim1.5''$ or modest improvements to $\sim0.1''$, respectively. While these missions can enable dramatic new scientific studies and goals \citep[e.g.,][]{axis}, they build off of an architecture that inherently limits our capabilities with X-ray imaging to observing larger scales, and resigning our window into the smallest scale processes to indirect spectral or polarimetric studies \citep[e.g.,][]{kim2024,saade2024,wilkins2026}. 

This technological gap represents a significant barrier towards truly understanding the most energetic processes in the Universe. For example, one of the most fundamental questions in extragalactic astrophysics is how supermassive black holes (SMBHs) grow to such immense sizes, up to billions of times the mass of our Sun. The primary method by which SMBHs grow is by accumulating material from its host galaxy into an accretion disk that ``feeds'' the central black hole. This process requires the gas to lose significant amounts of angular momentum in order to reach the SMBH, however the mechanisms by which this happens remains largely unclear as the accretion disk size for even the most nearby X-ray active SMBHs is just $\sim50~\mu$as in diameter \citep{eht2019,Lu2023}, far below the resolution capabilities of current or proposed X-ray facilities. Studying the high energy emission from the accretion disk will give us a window into how this angular momentum is transferred from the disk, and therefore how the SMBH feeds.

Closer to home, extreme stellar flares can significantly affect planetary atmospheres and environments relevant to habitability on a planet, either by promoting prebiotic chemical processes or by eroding planetary atmospheres that are important for maintaining habitable surface conditions \citep[see][for reviews]{AlvaradoGomez2022_CMEs_exoplanets,segura2026}. To date, the only resolved images we have of these flares are for our own Sun, with resolutions as fine as tens of km \citep[e.g.,][]{tamburri2025}. A resolved picture of the spatial distribution of flares across a larger sample of stars would offer critical insight into active region dynamics and their effects on planetary atmospheres, with implications for habitability.

Via a NASA Innovative Advanced Concepts (NIAC) study, we have examined the Accretion Explorer Interferometer (AEI), a mission concept for X-ray imaging that could achieve up to six orders of magnitude better resolving power than the current state of the art \citep[][]{weaver2026b}. AEI is designed to resolve black hole accretion and image stellar coronal structures and flaring activity. 
A detailed discussion of the AEI architecture can be found in the NIAC report \citep{weaver2026b}. In Section \ref{sec:AEI}, we briefly describe the concept and key technical specifications. 
With its architecture, AEI attains $\mu$as resolution with an effective area that is currently designed to approximate the current Chandra ACIS effective area of $\sim100$~cm$^{2}$ at 1.0 and 6.4 keV\footnote{\url{https://cxc.cfa.harvard.edu/caldb/prop\_plan/pimms/index27.html}}.

This dramatic increase in spatial resolution opens up a fundamentally new discovery space, enabling science cases to directly peer into the innermost regions of an accreting SMBH, from the fundamental mechanics of accretion to the launching of powerful jets that drive energy into the larger galaxy \citep[see][for a more complete review]{weaver2026}. 
Our team examined four science pillars to assess the most transformative science cases for an X-ray interferometer reaching $\mu$as resolution. The high-level science traceability matrix (STM) is published in our NIAC report \citep{weaver2026b}. Here we report additional details for the science cases investigated in the NIAC study, and expand to include other science priorities for an ultra-high resolution X-ray imager. Finally, in Section \ref{sec:STM}, we identify open questions regarding the new scientific discovery space within reach with a $\mu$as-scale X-ray interferometer and highlight places where precursor science could provide further insight into these future science goals.

\section{The Accretion Explorer Interferometer (AEI)}\label{sec:AEI}

\begin{figure}[ht!]
\centering
  \begin{subfigure}[b]{0.95\textwidth}
 \includegraphics[width=\textwidth]{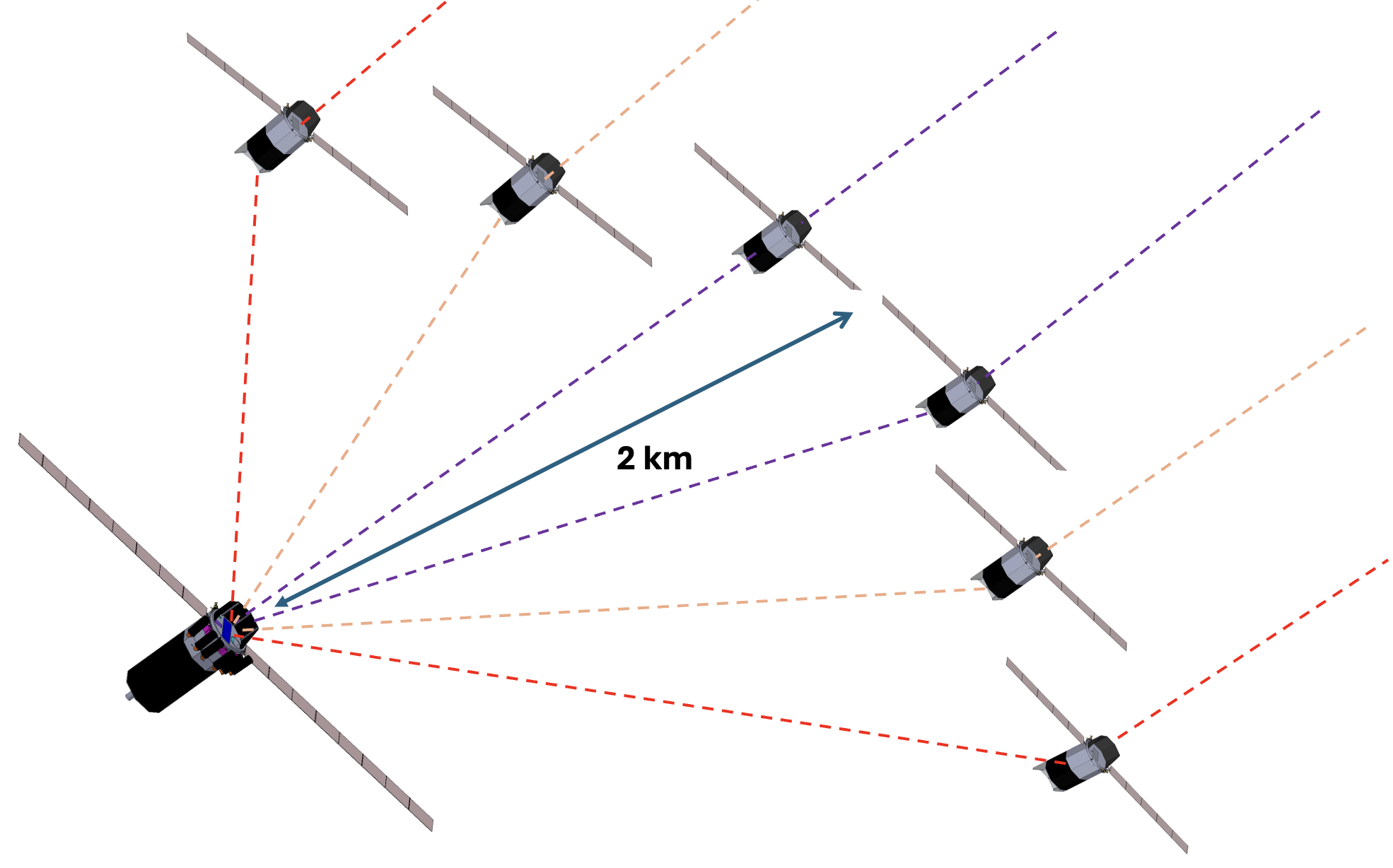}
     \caption{Deployed AEI array at L2. The satellite that holds the detectors and combiner optic is to the lower left and the other six smaller satellites contain mirror flats to collect and redirect X-rays from the target (indicated by dashed lines). Array curvature is exaggerated for display.}
  \end{subfigure}
  
\bigskip

  \begin{subfigure}[b]{0.95\textwidth}
\includegraphics[width=\textwidth]{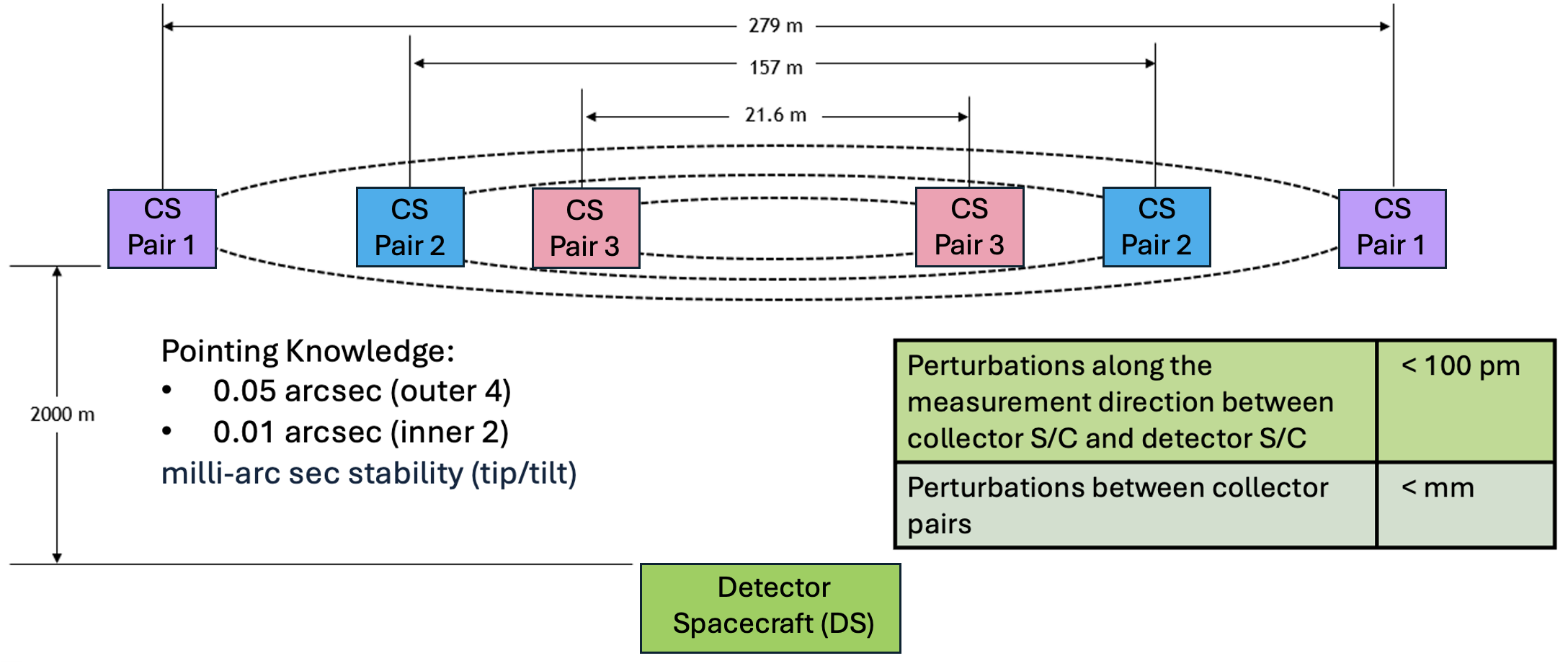} 
    \caption{Array alignment chart.}
  \end{subfigure}
    \caption{\small {\bf The AEI operational array configuration.} (a) The conceptual idea of the deployed array showing the light paths from collectors to the detector. (b) a visual chart showing all of the relative distances between the spacecraft, including baselines, and listing the needed stabilities for maintaining the required pathlength differences that would allow the configuration to maintain phase information and produce X-ray fringes.
}
   \label{fig:constellation} 
\end{figure}

The AEI concept consists of seven satellites flying in formation that build the interferometer, with one ``detector" satellite and three pairs of ``collector satellites" flying approximately in line $\sim2$ km in front of the detectors (Figure \ref{fig:constellation}). Each pair of collector satellites forms its own baseline with specific pair separations of 21.6 m (inner pair), 157 m, and 279 m (outer pair). The set of primary mirrors on each collector pair reflects X-rays at grazing angles, directing them to a secondary combiner optic on the detector satellite. The detector satellite hosts two X-ray spotting telescopes that operate independently, plus the combiner optic for the $\mu$as interferometer that is made of an X-ray beam splitter (see Rohrbach et al., in prep for more details). This optic potentially eliminates the need for long focal lengths \citep{cash2003a,cash2005imaging} that have historically proven a barrier towards mission feasibility for X-ray interferometry. The collector satellites are arranged to probe key energy bands required by the proposed science cases: $\sim0.7-1.2$~keV, $\sim1.5-2.5$~keV, $\sim6-7$~keV. Energy band coverage could be expanded by adding more collector pair baselines. The detectors are traditional large-format X-ray CCDs, which set the energy resolution and narrow-band imaging range of the interferometer to be E/$\Delta$E $\approx30-50$.

\section{Science Case 1: An Unprecedented Look into the Central Engine of an AGN}

\subsection{Understanding the Mechanisms of Accretion}

The accretion of material onto a SMBH is almost always associated with X-ray emission of up to hundreds of keV. These accreting supermassive black holes, known as active galactic nuclei (AGN), not only contain an accretion disk but are also surrounded by gas and dust where X-ray and UV photons from the disk can be scattered, absorbed, and/or reflected \citep[see][for a review]{netzer2013, beckmann2012}. This distribution of material is a complex interplay of inflows acting to feed the SMBH and outflows feeding energy back into the circumnuclear environment \citep{H&R2024review}. Currently, our models of SMBH growth and how they interact with their environment are oversimplified leading to possibly an incorrect picture of black hole and galaxy growth. Hard X-rays associated with most AGN are thought to originate from photons scattered by a so-called optically thin hot “corona” consisting of $\sim 10^{9}\,$K  electrons located at the innermost part of the disk \citep[e.g.,][]{haardt1991}. It is still an outstanding question how the electrons reached this temperature in the first place. The best guess at this point is magnetic reconnection, and, in fact, recent observations of X-ray flares in NGC 3783 using XRISM- and XMM-Newton-identified flares appear to agree with this hypothesis \citep{vardi2026}. However, this cannot be the only heating source as the material would cool too quickly to sustain the observed emission \citep{beloborodov2017}. Directly imaging the geometry of the corona could unlock the answer. Suggested coronal geometries range from a point source located above the SMBH to a ``sandwich” model that sits above and below the accretion disk \citep{bambi2021}. The corona appears to be located within a few tens of gravitational radii \citep{dai2010,kara2016} but a lack of spatially resolved imaging prevents us from directly confirming its nature. There is also evidence for a corona that expands and contracts, and this could depend on the SMBH mass and accretion rate \citep[e.g.,][]{wilkins2026}. For a SMBH of 100 million solar masses, the gravitational radius ($R_g$) would be $\sim300$ million km ($1\times10^{-5}$ pc). With AEI, we could observe within $\sim10~r_g$ for these types of AGN out to several tens of Mpc.

\begin{figure}[t]
\centering
 \includegraphics[width=0.9\textwidth]{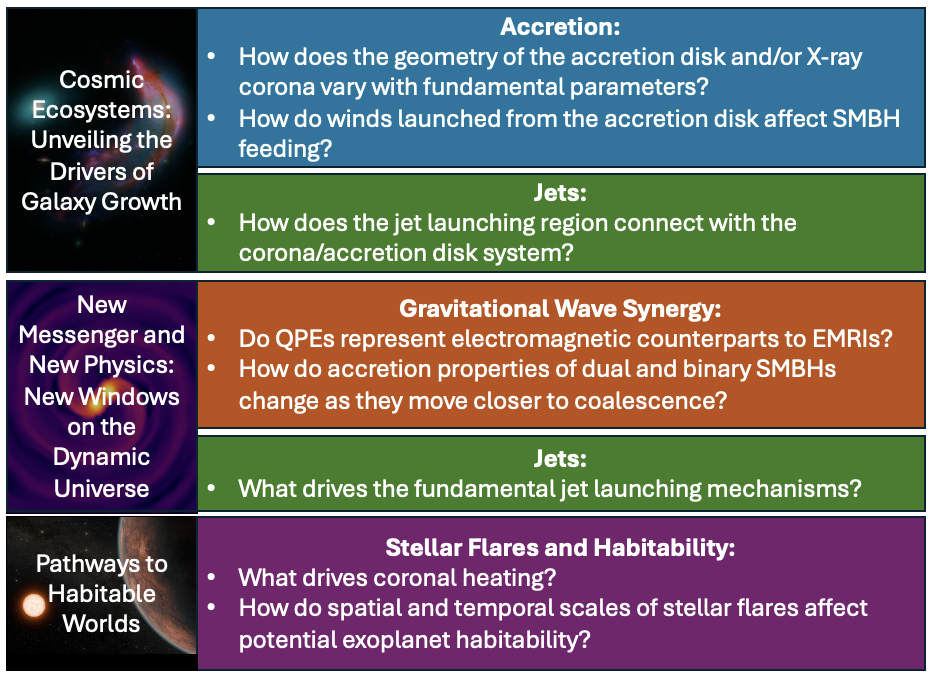}
\caption{\textbf{Science questions to be explored with AEI}. Above displays a subset of key science questions inspiring the science traceability matrix, connected with their respective Astro2020 \citep{Astro2020Decadal} Decadal Priority.
}
\label{Fig:accretion-questions}
\end{figure}

A considerable portion of X-ray emission in AGN is also reflected by the disk, producing signatures such as the iron K$\alpha$ emission line at 6.4\,keV. However, the disk can change shape depending on the accretion state \citep{liu2022}. A leap forward in our understanding of the ways black holes produce energy requires discriminating between disk models. Additionally, radiation or magnetic pressure can drive significant amounts of material from the disk in the form of winds \citep[reaching up to $40\%$ of the speed of light;][]{tombesi2014} which reflect significant portions of the K$\alpha$ line, potentially quenching the growth of the SMBH and disrupting the material farther out from the disk. Images of the X-ray emission line (disk) emission compared to the continuum emission and wind emission are vital to a full understanding of how black holes grow, evolve, and interact with their environment.

The ability to spatially resolve X-ray emission can also distinguish between true AGN activity and X-rays from compact extreme star-forming regions. This is especially true for recent astonishing objects found by the James Webb Space Telescope (JWST) – extremely compact galaxies at high redshifts where it seems almost impossible not to have a SMBH, but where their X-ray evidence is currently lacking \citep{ananna2024}. AEI would also enable tests of General Relativity (GR) and plasma physics in the strong gravity limit. Measuring the position of gas blobs entering a black hole as a function of time would be a sensitive way to discriminate between GR models \citep{caiazzo2019} and understanding the coronal geometry will tell us how such plasmas are heated to billions of degrees Kelvin \citep{Wilkins2016}. 

\subsection{Current Understanding of the Central Engine of an AGN}
With current and planned observatories, we do not have the resolution to probe these regions through X-ray imaging, and instead rely solely on indirect information from spectra or variability studies. These rely on assumptions about the physical conditions and geometry of the various regions, and moreover there are often degeneracies baked into those assumptions, resulting in a severely incomplete picture. 

The bulk of study of the X-ray corona has been done using X-ray reflection spectroscopy and polarization studies, due to insufficient angular resolution to image this region directly. These studies, supported by microlensing and reverberation mapping, have suggested that the corona is compact \citep[$\sim10$ $R_g$ from the central BH; e.g., ][]{kara2016} and may also be vertically elongated \citep[e.g., ][]{hancock2023}. To explain these observations, several models of coronal geometry have been proposed. However, recent observations from Imaging X-ray Polarimetry Explorer (IXPE) of unobscured AGN have shown significant extent of the corona along the direction of the disk, even when a jet is present  \citep[e.g.,][]{Gianolli2023}.

The size and geometry of the accretion disk is most commonly studied through reverberation mapping \citep[e.g.,][]{2021iSci...24j2557C}. The most common assumption for accretion disk geometry is the optically thick, geometrically thin Shakura-Sunyaev model \citep[][]{shakura-sunyaev1973}, however this model has been known to fail in cases with extremely high \citep{kubota-done2019} and extremely low accretion \citep{yuan-narayan2014} rates. Moreover, even in the moderately accreting regime, the standard disk model does not adequately reproduce the emission seen in observations \citep{koratkar1999,davis2011}. Accretion disk winds are generally studied through high-resolution spectroscopy of the Fe\,K$\alpha$ emission line \citep[e.g.,][]{Pounds2003}. The fastest winds can reach velocities up to 40$\%$ the speed of light and are thought to be launched from $\sim 0.0003-0.03\,$pc from the SMBH due to either radiation pressure or magnetic pressure \citep[e.g.,][]{Tombesi2012, Miller2025}] with slower winds launched from the outer edges of the disk \citep[e.g.,][]{tombesi2013}. However, significant uncertainties remain in the connection between the formation of the corona and the presence of a fast wind \citep[e.g.,][]{Fukumura2019}, as well as their exact launching location \citep[e.g.,][]{Laurenti2021}.

Finally, only a small number of AGN tori have been directly imaged, relying on interferometric observations from the Very Large Telescope Interferometer (VLTI) or the Very Large Array (VLA). The most famous example is NGC 1068, whose torus has been directly imaged by ALMA \citep{garcia-burillo2019}, VLTI/GRAVITY \citep{vermot2021}, and MIDI \citep{jaffe2004}, which provides insight into the size of the torus and its dust composition. Mid-infrared interferometric observations of other nearby galaxies such as Circinus \citep{tristram2014} and ESO 323-G77 \citep{leftley2018}, as well as larger samples \citep{asmus2016} have shown that a significant fraction of the infrared emission comes from an extended polar wind, with a much smaller ``torus" component. Multi-wavelength observations of AGN tori have identified a toroidal “elongated ring” structure, or multiple stratified rings, circling the central black hole, with diameters of only a few parsecs long, however there have been no comparable resolution X-ray observations.

\begin{figure}[t]
\centering
\includegraphics[width=0.9\textwidth]{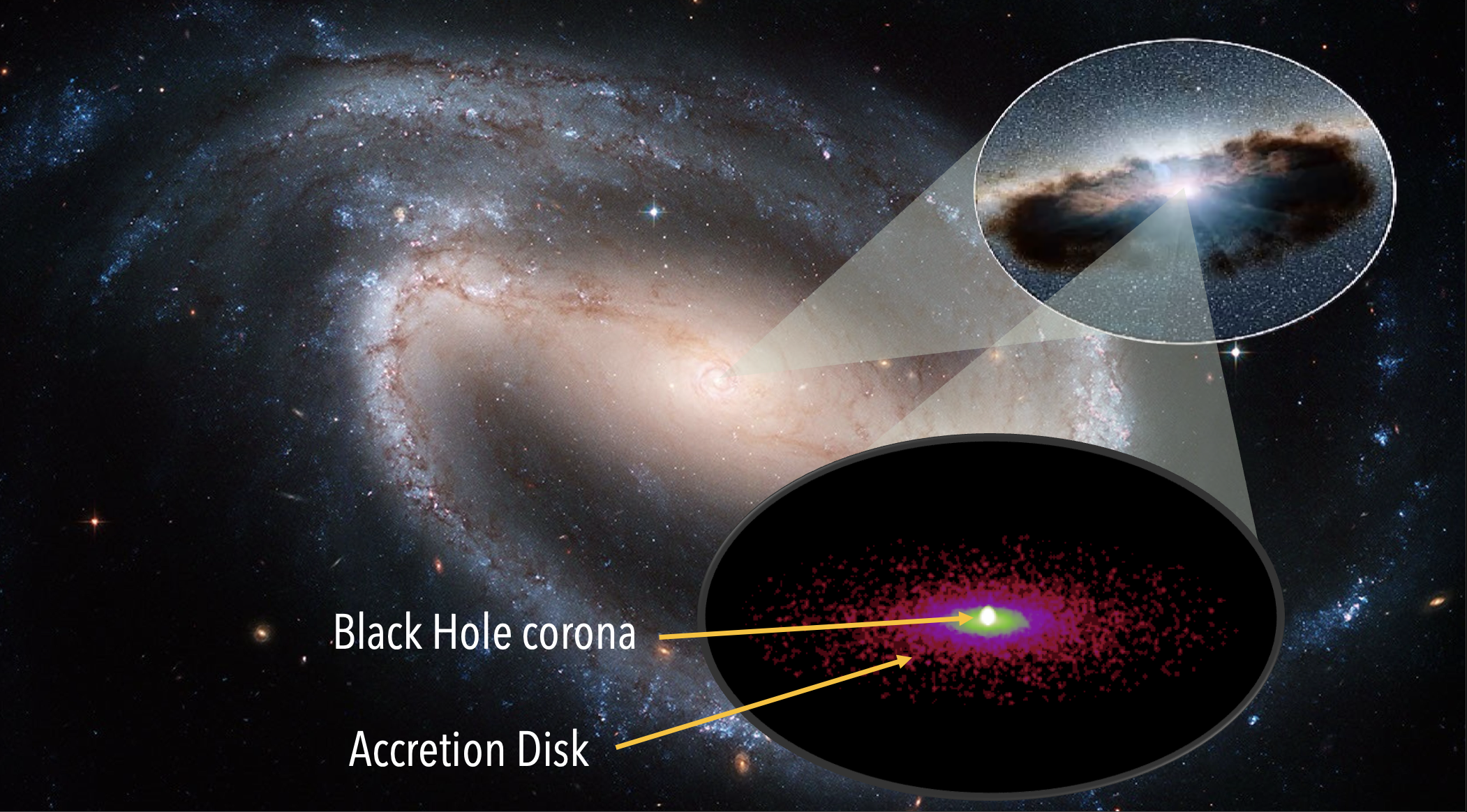}
\vspace{0.3cm}
\caption{\textbf{Simulated image of an AGN accretion disk and hard X-ray corona observed with $\mu$as resolution.} The simulated disk was made assuming an accreting $10^{8}~\mathrm{M}_{\odot}$ SMBH at a distance of 5\,Mpc from Earth, consistent with the closest objects where we can spatially resolve the corona with $\mu$as resolution. Colors indicate intensity, with white indicating the highest intensity. This spherical corona has a radius of $10~r_{\rm g}$, in accordance with recent X-ray reverberation studies \citep{kara2016}. 
}
\label{Fig:corona_models}
\end{figure}

\subsection{Technical Requirements}

The most stringent capabilities required for the successful completion of this science case are angular resolution and sensitivity. Even in the nearest X-ray active SMBHs, the X-ray corona only subtends $\sim\mu$as-scale angles. The angular resolution needed to directly image the accretion disk is similarly stringent, requiring resolutions of $10$s of $\mu$as to resolve $\sim5$ accretion disks of local AGN, and $\mu$as resolution for larger samples of up to $\sim20-30$ disks.

Of equal concern is the sensitivity needed to produce an adequate picture at these extreme resolutions, however there is significant uncertainty in these calculations, as AEI will probe resolutions up to six orders of magnitude beyond current X-ray images. We therefore use a variety of methods to calculate a range of potential sensitivity requirements. First, according to Uttley et al. 2021 \citep{Uttley2021}, 2-10 keV fluxes for a detailed study of inner regions of the nearest powerful AGN would be $\sim10^{-11}$ erg~cm$^{-2}$~s$^{-1}$ (for 100$\mu$as resolution).
Using a rough linear scaling factor ($10^4$) to scale that flux down to the $\sim\mu$as value, we find fluxes of $10^{-15}$~erg~cm$^{-2}$~s$^{-1}$. However, it is probable that the bulk of the X-ray emission will be concentrated within small spatial scales of the corona-disk combo, and therefore that the true sensitivity needed to resolve the brightest nuclear regions could be $10^{-13}-10^{-14}$~erg~cm$^{-2}$~s$^{-1}$ or higher. 

For a second estimate, we use the Fe K$\alpha$ ($\sim2-3\times10^{-13}$~erg~cm$^{-2}$~s$^{-1}$) and 2-10 keV ($\sim10^{-11}$~erg~cm$^{-2}$~s$^{-1}$) fluxes from a reverberation mapping study of NGC 3516 \citep{noda2023}. As the narrow component of Fe K$\alpha$ is believed to be mostly within $\sim10$ pc ($\sim$mas at $z\sim$0.009), we roughly linearly scale this down to $\sim10^{-15}$~erg~cm$^{-2}$~s$^{-1}$ for 2-10 keV and $\sim10^{-17}$~erg~cm$^{-2}$~s$^{-1}$ for the Fe K$\alpha$ emission line. 
Finally, we calculate sensitivity requirements based on a rough scaling from Chandra data. Taking the Fe K fluxes and predicted extents from Andonie et al. 2022 \citep{andonie2022} and linearly scaling to $\mu$as level, we predict fluxes of $\sim10^{-15}$ erg~cm$^{-2}$~s$^{-1}$ - $\sim10^{-16}$ erg~cm$^{-2}$~s$^{-1}$. We emphasize that linear scaling is very likely a significant oversimplification, and that the flux sensitivities presented here should therefore be considered as stringent lower limits, with the true required sensitivities potentially several orders of magnitude higher.

As discussed in Weaver et al. 2026 \citep{weaver2026} and Uttley et al. 2021 \citep{Uttley2021}, imaging the accretion disk system requires a field of view of at least $100\mu$as, well within the $\sim400\mu$as field of view considered for the current architecture of AEI. If a trade is needed to improve sensitivity or resolution, however, this constraint could be relaxed in favor of mosaicking capabilities with precise pointing and calibration to ensure the effective combination of images.

The proposed AEI hard band ($\sim6-7$~keV) would be sufficient to probe Fe K emission from the corona, though access to harder continuum would better constrain separate emission from Fe K reprocessing as opposed to continuum seen directly from the corona. The proposed softer bands ($\lesssim 2.5$~keV) would probe feedback and winds come from the accretion disk and, in concert with the harder bands, explore obscuration within the central engine. Finally, CCD-quality spectral resolution is sufficient and would allow for the direct study of prominent spectral features in sources with sufficient counts.

\section{Science Case 2: A High-Resolution Look into Sources of Gravitational Waves}



SMBHs can also power some of the most dynamic high energy events in the Universe. In the advent of the Laser Interferometer Space Array (LISA) gravitational wave (GW) observatory, there is significant interest in improving our understanding of the electromagnetic counterparts of GW events associated with periods of SMBH growth. Our understanding of the growth patterns, dynamical evolution, and the transition phase between dual SMBHs and binary SMBHs between $10-100$~pc (and hence the pairing rate of binary SMBHs) is now largely informed by large scale cosmological simulations \citep[e.g.,][]{kelly2017,chen2023}; no dual AGNs have been confirmed with separations 30 pc $< r_p <$ 200 pc to anchor the theoretical constraints, and only a single dual AGN with separation 200 $< r_p <$ 400 pc has ever been confirmed. Meanwhile, our understanding of the accretion disk physics, structure, and properties of sub-pc-scale binary SMBHs are entirely driven by theoretical modeling \citep[e.g., ][]{noble2021}. Moreover, while current instrumentation has identified semi-periodic episodes of accretion of a compact object onto a supermassive black hole 
\citep{kara2025}, our window into this extreme episodic growth remains indirect, and primarily via spectral and timing data.

\subsection{Selection and Study of Binary Candidates via Direct Imaging}
Currently, the only way to directly resolve pc- and sub-pc scale binary systems at typically observed redshifts ($z > 0.01$) is via radio very long baseline interferometry (VLBI) which can achieve resolutions as low as $\sim0.1$~mas. However, the vast majority of AGNs are not expected to be radio bright, so this method thus biases us against the majority of the population. Also, current and planned radio VLBI facilities cannot achieve resolutions necessary to resolve binary AGNs in the continuous wave regime ($\lesssim10^{-3}$ pc). We are currently limited to purely theoretical constraints for the dynamical and mass accretion evolution of these SMBH pair populations.


As X-ray emission is a ubiquitous feature of AGNs, $\mu$as resolution X-ray imaging would have a dramatic impact on our understanding of binary accretion activity, accretion disk structure, and angular momentum losses at such separations, which would be impossible at lower angular resolutions. Furthermore, as milliparsec and sub-milliparsec binaries fall within the continuous gravitational wave regime \citep[e.g., ][]{burke-spolaor2018}, $\mu$as X-ray imaging and pulsar timing arrays (PTAs) or space-based gravitational wave detectors such as LISA could offer an unparalleled synergy capable of resolving distinct continuous wave sources and understanding the interplay between accretion activity, SMBH growth, and gravitational wave emission in these inspiralling binary systems. 

CSO 0402+379 \citep[$z=0.05$, separation $\sim7~$pc][]{rodriguez2006} is the only currently known binary SMBH and was originally detected (serendipitously) using radio VLBI. The fact that both nuclei are active (i.e., radio-bright) suggests ongoing accretion and implies dissipation of the angular momentum of the system. This system has been observed in the X-rays \citep[e.g.,][]{andrade-santos2016,mondal2024}  with clear Fe K emission indicative of ongoing accretion, making it a perfect target for X-ray interferometry.

\subsection{QPEs - a possible sign of EMRIs?}
Quasi-periodic eruptions (QPEs) are periodic ($\sim$hour timescales) and extreme bursts of luminosity that originate from a galaxy's central SMBH. Due to spatial resolution limitations, our current understanding of QPEs is primarily relegated to timing studies \citep[see][for a review]{kara2025}. 
The serendipitous discovery of the first of these sources \citep{Miniutti2019} with XMM-Newton ushered in a new era of time domain astronomy, in an attempt to identify the origin of these mysterious flares. Since then, over ten have been discovered \citep[see][and references therein]{goodwin2025}, all showing similar soft X-ray spectra and recurring behavior. Most of these sources occur in galaxies with known tidal disruption events (TDEs) or changing look AGN (CLAGN), suggesting that this emission can be tied to a dramatic change in accretion flow \citep[e.g.,][]{Miniutti2019,chakraborty2021, quintin2023, nicholl2024}. This is further suggested by spectral analyses, where the spectrum during a flare shows a much hotter ($kT \sim 100-250$ eV vs. $kT \sim50-80$ eV) thermal component than in a non-flaring period \citep{kara2025}. Current models suggest that these events could be caused by accretion disk instabilities and/or an extreme mass ratio inspiral (EMRI), however additional study is needed to confirm their origins \citep{kara2025}. EMRIs are particular sources of interest for LISA due to the insight and constraints they can provide on black hole spin and the accretion history of the SMBH. While the detection of a QPE correlated with a gravitational wave signal from LISA would provide robust evidence towards the EMRI hypothesis, EMRI detections with LISA are expected to be extremely complex, and sky localization capabilities may be limited \citep{amaroseoane2023}. High spatial resolution X-ray imaging accessible with Accretion Explorer could provide a more direct look into these sources, offering additional insight into the physics driving this flaring behavior. 
To move beyond a dependency on timing studies, a facility capable of resolving SMBH pairs down to sub-pc separations, and directly tracing the interaction-driven accretion events believed to be the source of QPEs, is a crucial step in understanding the physics, structure, and properties of accretion flows for merger-driven SMBH growth.




\subsection{Technical Requirements}

The most critical requirements for this science case are sensitivity and angular resolution, with a strong need to resolve close pair separation systems or resolve the accretion disk system with sufficient counts for robust detection. Note that if one goal within this science case (binary candidate, QPE) provides a more stringent constraint than the other, we present the most stringent request. 

First, $\mu$as resolution would enable us to resolve milliparsec (mpc) scale binaries throughout the universe, including sub-mpc ($<10^{-3}$ pc) binaries at z $\lesssim0.05$. Moreover, as discussed above, $\mu$as resolution would enable the resolution of the accretion disk, providing the opportunity to directly observe the mechanisms behind QPE emission. A field of view of $\sim400$~$\mu$as is optimal to image the full extent of compact binary AGN systems (e.g., SMBHs and associated accretion disks and mini disks, if present), with milliparsec ($1\times10^{-3}$ pc) binaries observable throughout the Universe and $<0.01$ pc and $<1$ pc binaries observable beyond redshifts of $z\approx0.001$ and $z\approx0.1$, respectively, without the need for mosaicking. However, mosaicking is acceptable and could increase the observable sample through observations of systems with larger separations -- such as binaries with separations of tens of parsecs or dual AGNs with separations of tens to a couple hundred parsecs -- as long as, if necessary, pointing and calibration are sufficient to precisely combine the images.

Similarly to the accretion science case, spectral coverage including a ``broadband" hard X-ray bandpass ($\gtrsim 2$~keV) and a softer ($\lesssim2$~keV) could enable the computation of hardness ratios and basic spectral shapes. A third filter covering the Fe K region ($\sim6-7$~keV) could be used to probe emission from the accretion disk. 
As discussed above, sensitivity estimates hold significant uncertainty at this stage.
We use the 2-10 keV and 6-7 keV luminosities of confirmed binary CSO 0402+739, and find them comparable to those investigated in the above accretion science case, so targets in the local Universe will have comparable sensitivity limits. For higher-$z$ sources, we scale luminosities of $10^{42}$~erg~cm$^{-2}$~s$^{-1}$ to $z\sim0.8$, which is the highest redshift for a sample of strong binary candidates in the Big Multi-AGN Catalog \citep[Big MAC;][]{pfeifle2025}. Assuming that at these distances, the X-ray sources would be primarily point sources with limited extended emission, we predict a sensitivity limit as low as $\sim10^{-16}$~erg~cm$^{-2}$~s$^{-1}$. 

CCD-quality spectral resolution would be optimal to provide direct constraints on X-ray spectroscopic properties, such as the intrinsic power law slope and normalization, line-of-sight column density, etc. If CCD quality is not attainable, a few spectral windows across the soft and hard bandpasses would be sufficient for emission and spectral mapping. Akin to identification of radio core and jet structures in radio VLBI imaging, this spectral resolution would enable pairs of X-ray AGNs to be robustly confirmed and star formation/radio jet-related X-ray emission to be ruled out as origins of the observed emission (whereas at lower energies, these sources may dominate if present).

The most stringent requirement for effective area comes from the need to detect moderate luminosity QPEs within potentially short flaring periods. For this, an effective area comparable to Chandra's ($\sim100$~cm$^{2}$) would be sufficient to detect moderate luminosity QPEs with significant counts ($\sim20$) within $<1$ hr exposures, which is required to probe the full cycle (e.g., before, during, and after the event). Exposure times of $\sim$days to a week or more are required for the highest resolution studies of binary SMBHs.

\section{Science Case 3: Understanding Stellar Activity and its Effects on Habitability}


The search for habitable environments beyond Earth has been a topic of scientific interest for decades, and has inspired the scientific goals behind the Habitable Worlds Observatory mission concept. As a result, the conversation around habitability has similarly become more nuanced, moving from rough calculations of a habitable ``Goldilocks zone'' to more intricate discussions of atmospheric escape and the larger host star--exoplanet connection \citep[see][for a review]{seager2013, kane2021}. The effects of stellar activity on a potentially habitable planet can be extreme; stellar flares and CMEs could expose planets to high-energy radiation driving atmospheric loss and altering surface radiation environments.

\subsection{The Dynamic Stellar Corona}
The stellar corona is an intricate system, with magnetic structures that regulate stellar activity and shape the surrounding space weather environment. Of particular note are magnetic coronal loops, which are arc-like structures of dense plasma confined within magnetic flux tubes. These structures are central to our understanding of stellar magnetic dynamos \citep{Gudel_2004_AARv, Reiners_2012_LRSP, Testav_2015_SSRv,Reale-04-A+A-2, Reale-07-A+A-2, Karmakar-16-MNRAS-9,Karmakar-17-ApJ-5}; however, they remain unresolved beyond the Solar System, severely limiting our understanding of space weather drivers in stars other than our Sun. 

The mechanisms behind coronal heating, and in particular how the corona reaches and sustains such high temperatures compared to the photosphere, are still largely uncertain \citep{Gudel_2004_AARv,Wood_2004_LRSP, ArgiroffiC-19-NatAs-1}. Nanoflares are a leading candidate mechanism for coronal heating, although their contribution relative to other heating mechanisms remains uncertain \citep{Favata-00-A+A-3,Osten-10-ApJ-5,KarmakarS-22-MNRAS-1,KarmakarS-23-MNRAS}. Direct X-ray imaging would provide a complementary test of Zeeman-Doppler Imaging (ZDI)-based coronal models by determining whether structures inferred from large-scale surface magnetic maps correspond to coherent coronal structures or unresolved ensembles of smaller magnetic features.




\subsection{Exospace Weather and Habitability}
High-energy stellar radiation acts as a double-edged sword for planetary evolution. While intense X-ray and UV fluxes can strip away atmospheres and threaten long-term habitability, they may also drive chemical pathways relevant to prebiotic chemistry \citep[e.g.][]{2007AsBio...7..185L,airapetian17,Maliniemi22_SEP_depleteOzoneEarth,airapetian21_adv_n_disadv_SEPsforlife}. Active M dwarfs and RS CVn binaries occupy the high-activity end of the stellar population, while relatively quiescent F and G stars sit at the lower end \citep{Freund_2018,Magaudda_2020,Mohan_2021_EMISSA}. Meanwhile, M dwarfs account for $\sim70\%$ of stars within the Milky Way and host most of the known close-in low-mass exoplanets in habitable zones, followed by K and G dwarfs~\citep{2020A&A...643A.106B}. Hence, understanding the effects of the magnetic activity of G-M dwarfs on planetary systems is important for assessing potential habitability.
The ability to resolve stellar coronae and localize the origins of flares at $\mu$as spatial resolution would provide an important advance, as solar studies show that CME occurrence varies with coronal latitude and the solar activity cycle, but CME latitude distributions remain essentially unconstrained for other stars \citep{Yashiro_2004_JGRA,Kane2008,Webb_2012_LRSP,Lin2022}. 
Rather than viewing low-mass stars as blurry point sources, $\mu$as imaging would allow us to spatially isolate individual flare footprints and map their geometries. Combined with ultra-fast temporal resolution (sub-minute to sub-second cadence), this would yield new observational constraints on how localized flare morphology affects orbiting planets.

This necessity for ultra-high temporal and spatial resolution is underscored by observations of our own Sun, where sub-second particle acceleration signatures are routinely captured in radio and X-ray maps of active regions. These rapid timescales are vital for distinguishing between thermal and non-thermal plasma components within the low corona and chromosphere \citep[e.g.][]{Morosan14_millisecLOFARimging,Mohan17_SubsecSolarvariability,Chen19_VLAMillisecImging,Mohan_2019,Mohan23_solar-stellarSnapshotimaging,Juan23_ALMASec-scaleChromosphere,skokic23_all_Mmflares_sun_multibandstudy}. Translating this capability to other stars at $\mu$as spatial resolution would allow us to map the latitudinal and spatial distributions of stellar flares, providing new physical constraints on active region dynamics. Such measurements would help constrain how extreme space weather and CME-driven atmospheric erosion influence the long-term evolution of exoplanet atmospheres. 

By leveraging $\mu$as resolution, we can extract high-cadence light curves from spatially resolved regions of the stellar disk, effectively disentangling overlapping flare events. This spatial isolation will allow for the calculation of spatially resolved flare frequency distributions (FFDs) for specific active regions on planet-hosting stars. Consequently, these measurements would constrain the high-energy radiation environments experienced by close-in planets and provide inputs for models of atmospheric escape and photochemistry \citep{Rimmer2018}.

\begin{figure}[htbp]
    \centering
    \includegraphics[width=\textwidth]{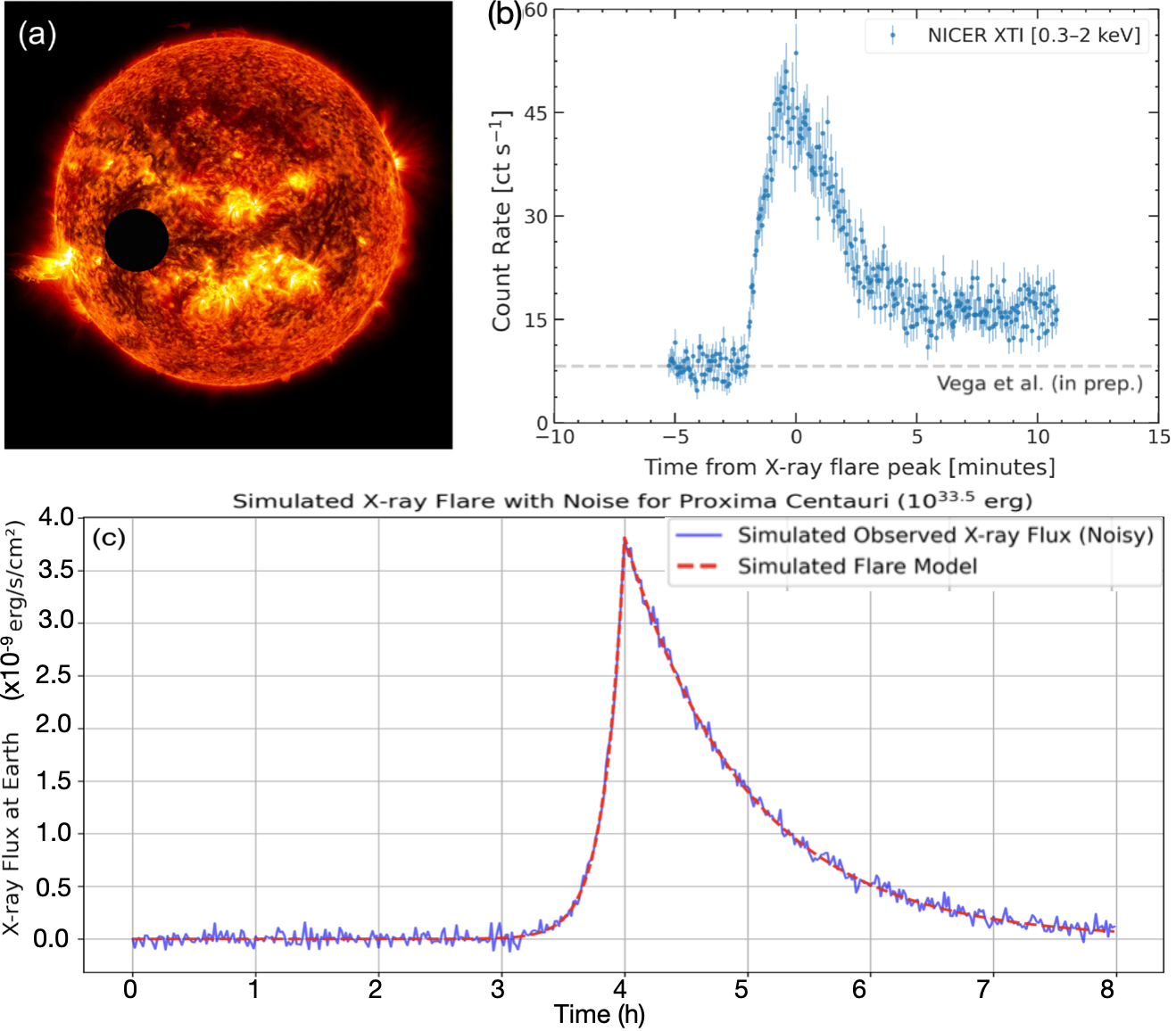}

    \caption{Proxima Centauri as a candidate for high-resolution X-ray observations with AEI:
    (a) A combined SDO/AIA 304 and 171~\AA\ solar image is shown as a template of an active corona of a cool main-sequence star (F-M type). An orbiting exoplanet is shown as a dark patch.
    (b) NICER 0.3--2~keV X-ray observations of a flare from the active M dwarf YZ CMi, with an estimated X-ray energy of $\sim10^{31.7}$~erg \citep{Vega2026}.
    (c) Illustrative simulated X-ray flare with an assumed energy of $\sim10^{33.5}$~erg. For comparison, Proxima Centauri is known to exhibit superflares with total energies reaching approximately this scale \citep{Howard2018}; the simulation is illustrative and does not imply an equivalent observed X-ray flare energy.
    Proxima Centauri's quiescent X-ray flux is approximately $2.6\times10^{-12}$~erg~cm$^{-2}$~s$^{-1}$ in the 0.2--10~keV band, corresponding to a luminosity of approximately $4.9\times10^{26}$~erg~s$^{-1}$. This level of X-ray emission is comparable to that of the solar corona during periods of low activity \citep{Fuhrmeister2011}.}
    \label{fig:prox_cen_images}
\end{figure}

\subsection{Technical Requirements}
The main technical priorities needed to achieve this case are spatial resolution, sensitivity, and time resolution in order to both detect and spatially isolate rapid transients such as flares. 
Typical solar active regions that power strong flares, coronal mass ejections (CMEs), and energetic particle (EP) events can vary within $\sim50 - 150$\,Mm~\citep{Ligget85_ARscale,seiden96_ARscale,Svanda25_ARscale}.  
Spatial resolutions of $\sim 1 - 100\,\mu$as FWHM are required to locate and spatially resolve such coronal active regions and probe large-scale erupting structures associated with massive CMEs in nearby stars located within 10\,pc. The scales mentioned are for the Sun, while most stars are M dwarfs with a radius of $\sim0.1 - 0.5$\,R$_\odot$, which means the expected activity scales can be an order of magnitude smaller. Hence, to sample active phenomena in a resolved manner and study the spatial scales of eruptive events across a sufficient sample of F-M dwarfs, a spatial resolution of the order of 10\,$\mu$as is preferable. For the nearby star, Proxima Centauri, spatial resolution of this scale will help explore even smaller-scale structures within active regions which are often associated with localised particle trapping and acceleration in coronal loops, often causing quasi-periodic pulsations (QPPs) in observed X-ray to radio band emissions. The spatial scales and temporal evolution of QPP sources can constrain local plasma and magnetic field conditions associated with particle acceleration in the corona, which are crucial in modeling the physics of EP production~\citep[see][for an overview]{2021SSRv..217...66Z,2026aask.bookE..19N}. 

The instrument must achieve a sensitivity of $10^{-12}$ to $10^{-13}$ erg s$^{-1}$ cm$^{-2}$. For example, a $5\sigma$ detection of Proxima Centauri's quiescent state demands $\sim 10^{-13}$ erg s$^{-1}$ cm$^{-2}$. While monitoring these faint quiescent states allows for time averaging on the order of minutes, timing resolutions of seconds to minutes are sufficient for most stellar-flare observations, with faster cadence desirable for resolving rapid impulsive phases.  
A larger effective area would improve photon statistics and enable higher-cadence imaging of quiescent stellar states at \(\mu\)as angular resolution.
Moreover, the detector architecture must handle highly localized flux variations of $\geq 3$ orders of magnitude (peaking up to $\sim 10^{-9}$ erg s$^{-1}$ cm$^{-2}$) without severe photon pile-up. This requires fast readout capabilities (e.g., sub-millisecond frame times) or advanced sensor arrays to preserve photometric and spectral integrity during extreme superflares \citep{vandenBesselaar2003_ADLeo_XMM_Chandra, Osten_2016_ApJ}.

Science targets can be selected to accommodate the 400$~\mu$as field of view afforded by this architecture. While full-disk imaging is possible for the most compact nearby targets, this narrow FOV facilitates a \textit{partial-disk} pointing strategy to track the emisison from identified active region latitudes directly in larger nearby stars (e.g., Alpha Cen, EV Lac, YZ CMi).

Broad soft-to-hard X-ray coverage is scientifically desirable to capture thermal coronal line emission (e.g., Fe, Ne, O), hot (\(>30\) MK) flare plasma, and nonthermal emission where present. CCD-quality resolution (\(R \sim 20-50\)) is the minimum required. However, moderate to high resolution (\(R \sim 100-1000\)) is highly desirable to map Doppler shifts, velocity fields, and isolate key highly ionized Fe lines, including Fe XXV near 6.7 keV and Fe XXVI near 6.97 keV \citep[e.g.][]{Gudel_2004_AARv, Osten2007}, along individual spatially resolved coronal structures. The current AEI architecture, sampling key bands near \(0.7-1.2\), \(1.5-2.5\), and \(6-7\) keV, is sufficient.

Finally, long  duration monitoring (e.g., spanning $1$ to $3$ days per target) is highly desired. While individual temporal bins will be minutes or less, spanning a multi-day window ensures a high probability of capturing extreme stochastic flare onsets and extended decay phases. This will help track the QPPs often seen in the pre-flare to post-flare phases carrying information about the varying plasma and magnetic field conditions in the active regions.
This strategy does not strictly require 100\% continuous staring; the monitoring can accommodate necessary mission operations (e.g., station-keeping maneuvers, telemetry) provided the overall temporal cadence remains dense.

\section{Science Case 4: Revealing Relativistic Jet Physics and Geometric Structure}


Relativistic jets from AGN represent a dynamic source of feedback whereby the central AGN can influence its host galaxy. Despite decades of study, significant open questions remain regarding the launching, acceleration and collimation of AGN jets, with limitations in current facilities and theoretical simulations preventing a robust study into the physics of these sources. Much of our observational study on jet physics has been using radio VLBI due to the high angular resolutions needed to resolve the inner regions \citep[see][for a review]{hardcastle2020}. X-ray studies of jets have typically focused on the large-scale structure \citep{Meyer2023} or polarimetry studies of the magnetic field geometry and particle acceleration mechanisms of blazars \citep[see][and references therein]{kim2024}. There are significant degeneracies in simulations, and open questions remain regarding the connection between the jet launching region and the innermost regions of the AGN.

X-ray studies of jets have made significant progress within the past few decades. Chandra's sub-arcsecond imaging remains the workhorse for resolved kpc-scale jets, mapping individual knots in prototypical sources like M87, 3C 273, and PKS 0637-752, where X-ray emission is co-spatial with radio knots in FR-I sources \citep{perlman2005,jester2006,schwartz2000}. This is consistent with synchrotron emission from electrons as power-law or $\kappa-$distributions cooling on $\sim$year timescales \citep{Hardcastle2016, WilsonYang2002, Marshall2002, Marshall2010}. In powerful FR-II quasar jets, the X-ray peaks are typically displaced from the radio, where IC/CMB scattering or distinct electron populations remain debated. \citep{Meyer2023} found that X-ray emissions of a statistically significant number of jets varied over just a few years, suggesting that the X-ray emission is produced from a region much smaller than the cross-section of the jet. More recently,  polarimetric observations of jets with IXPE have offered new insight into the magnetic field geometry and particle acceleration mechanisms within AGN jets \citep[see][for a review]{kim2024}. However, our understanding of these sources is limited to an indirect inference from multi-wavelength or multi-instrument observations over a far larger area than the regions from which these jets launch.
The observation of these physical properties and their associated topological magnetic field features can be achieved by AEI, with $\mu$as resolution.

\subsection{Jet launching mechanisms}

Accurate and concise launching mechanisms of relativistic jets in AGN persists as an open question, with three theoretical frameworks dominating the discussion. The Blandford-Znajek (BZ) process \citep{BlandfordZnajek1977} extracts rotational energy directly from a spinning black hole through frame-dragged magnetic field lines threading the event horizon. This generates an outgoing Poynting flux that powers the jet. The available jet power scales as $P_{BZ} \propto a^2 B_{H}^2 r_g^2 c$, where $a$ is the dimensionless black hole spin and $B_H$ is the magnetic field strength threading the horizon, making this mechanism prominent for rapidly rotating black holes in magnetically arrested disk (MAD) configurations \citep{Tchekhovskoy2011}. The Blandford-Payne (BP) process \citep{BlandfordPayne1982} instead extracts rotational energy from the accretion disk through magnetic field lines anchored to the disk with material driven by centrifugal forces along field lines inclined at angles $\lesssim 60^\circ$ to the disk plane. At the inner regions $\lesssim 100\,r_g$, the BZ-driven jet spine and BP-driven disk wind likely coexist, producing a stratified outflow with the high-Lorentz-factor core surrounded by a slower magnetized sheath \citep{McKinney2006}. Both the BZ and BP processes have implied assumptions under force-free MHD for the current density needed to support observed acceleration descriptions. They both lack a self-consistent prescription for where the charge carriers are seeded from.

The third competing mechanism is the Penrose-Williams (PW) mechanism \citep{Williams1995, Williams2004} that extends the original Penrose process, in which negative-energy particles plunge into the ergosphere. The PW mechanism is positioned in a collective MHD regime where magnetic reconnection in the ergosphere allows magnetic field lines to extract rotational energy through the negative-energy states accessible only within the static limit. Recent GRMHD simulations have shown that ergospheric reconnection events can power transient jet emission and may contribute alongside steady BZ outflows during magnetic flux eruptions in MAD states \citep{Bransgrove2021, Ripperda2022}. Distinguishing between these mechanisms requires direct imaging of the jet launching region at scales $\lesssim 100\,r_g$, where the magnetic field topology, disk-jet geometry, and ergospheric dynamics become observationally accessible. $\mu$as X-ray imaging of the corona-jet interface region would directly probe where the particle supply lives and how it connects to the launching region, which is precisely the gap in the analytical formalism.

\begin{figure}[t!]
  \centering
  \includegraphics[width=0.9\textwidth]{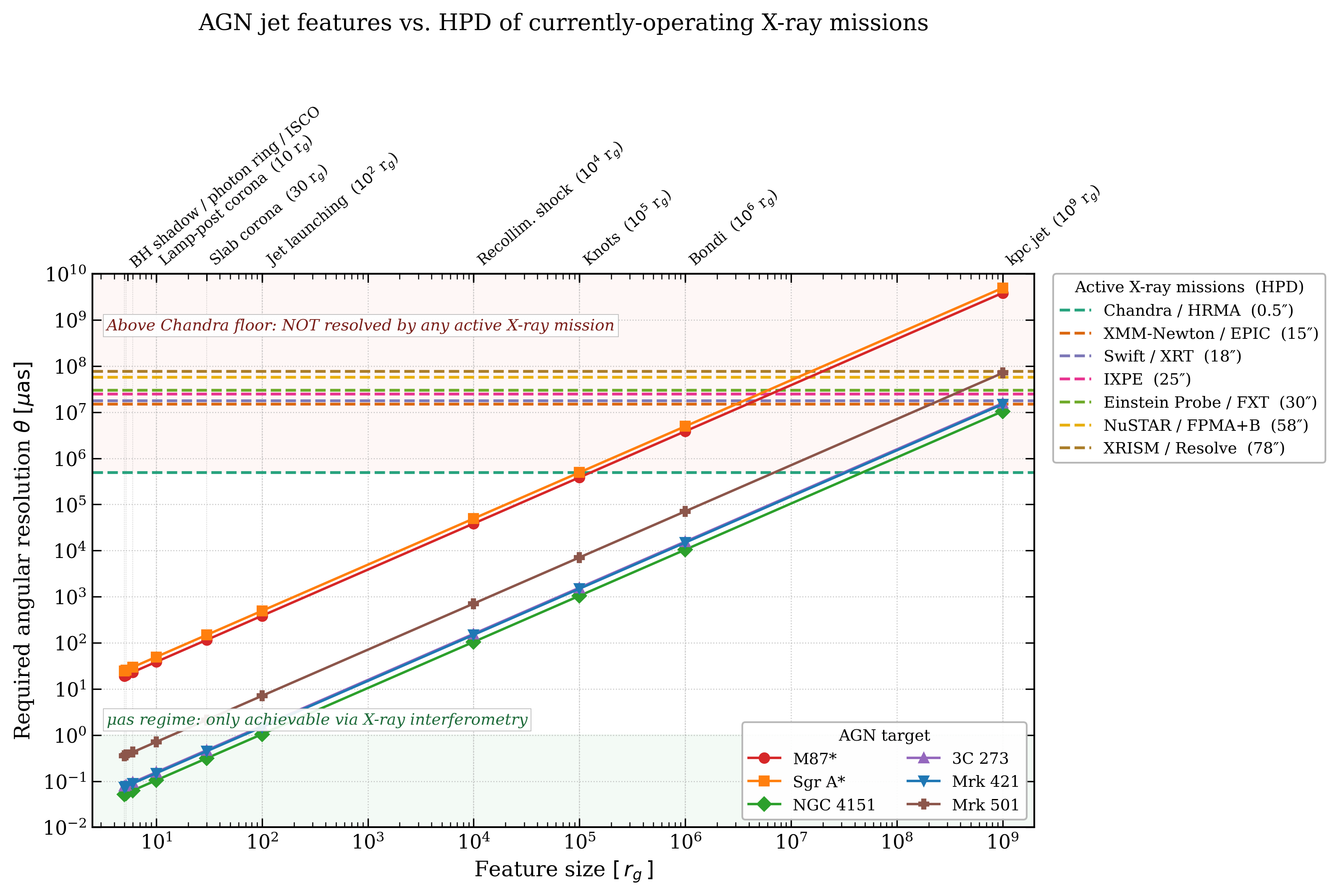}
    \caption{Angular size $\theta$ ($\mu as$) required to resolve each AGN jet feature for six different targets. Values are computed from the angular diameter distance $D_A = D_L/(1+z)^{2}$ and assume the characteristic feature scales of Table \ref{tab:feature_sizes}.}
     \label{fig:jetfeature_res}
\end{figure}

\subsection{Magnetic Field Topology and Polarization}

The \citep{Lobanov1998} method, with updated formulations from the MOJAVE dataset \citep{Pushkarev2012,Sokolovsky2011}, gives you the magnetic field strength as a function of distance from the jet apex, under the assumption of equipartition between particles and field. It does not give definitive information about the angular dependence in field geometry. That is whether the field generally supports helical, toroidal, or poloidal configurations. The synchrotron self-absorption that produces the core-shift depends on $|B|^{(p+2)/(p+4)}$ (with p the electron index), which is sensitive to magnitude but not orientation. Thus the VLBI core-shift gives you the radial profile $B(r)$ and the absolute distance scale in which the BZ jet power and MAD flux comparisons depend on. Polarization and Faraday conversion gradients would provide the necessary information for the field orientation.

A major shift has come from IXPE \citep{Weisskopf2022}. Since 2021, IXPE has detected significant X-ray polarization in more than a dozen high-synchrotron-peaked blazars (e.g. Mrk 421, Mrk 501, PG 1553+113, PKS 2155-304, 1ES 0229+200, 1ES 1959+650, and H 1426+428). This establishes a consistent pattern in which the polarization angle aligns with the jet axis and the wavelength dependence favors energy-stratified electron populations accelerated at compact shock fronts with locally ordered magnetic fields \citep{Liodakis2022, Ehlert2023, Middei2023, Errando2024, Kouch2024, Hodgson2025}.

\begin{table}[ht!]
  \centering
  \begin{tabularx}{\linewidth}{llll}
    \hline
    \hline
    Jet Feature & Size $(\sim r_g)$ & Baseline Criteria & Method \\
    \hline
    BH shadow            & $5$       & $\sqrt{27}\approx 5.20\,r_g$ (Schwarzschild) & Theory, Simulation \\
    Photon ring          & $5.2$     & $n=1$ photon ring & Theory, Simulation \\
    ISCO                 & $6$       & $\sim 1.2\,r_g$ for maximal Kerr & Theory \\
    Lamp-post corona     & $5-15$      & Reverberation at $h \approx 5$--$15\,r_g$ & Theory\\
    Slab corona          & $30$      & IXPE polarimetry & Observation (inferred) \\
    Jet-launching        & $50-200$  & M87 GRMHD collimation & Simulation \\
    Recollimation shock  & $10^{3}$--$10^{5}$  & M87 HST-1 & Observation (inferred) \\
    Inner jet knots      & $10^{5}$  & VLBI moving features & Observation (radio) \\
    Bondi radius         & $10^{5}$--$10^{6}$  & M87 HST-1 transition & Observation (inferred) \\
    kpc-scale jet        & $10^{9}$  & $1\,\mathrm{kpc}/r_g$ for $M_\bullet \sim 10^{9}\,M_\odot$ & Observation (X-ray) \\
    \hline
  \end{tabularx}
    \caption{Characteristic length scales of AGN jet features, in units of the gravitational radii $r_g \equiv GM_{BH}/c^{2}$ in geometric units for $G=c=1$. Here the baseline criteria lists standard signatures that derive and/or infer the corresponding jet feature.}
  \label{tab:feature_sizes}
\end{table}

\subsection{Technical Requirements}
The most critical priorities to ensure that this science case would be successfully explored are fine sensitivity, to probe faint features, and $\mu$as or better angular resolution, which would offer the opportunity to directly observe the jet launching region in the X-rays for the first time.

In particular, this case requires sensitivities of $\sim 10^{-15}$ erg cm$^{-2}$ s$^{-1}$ in a $10$~ks integration at 1 keV (point-source limit), driven by the requirement to fringe-track on M87's nucleus ($F_{2-10\,\mathrm{keV}} \sim 4 \times 10^{-13}$ erg cm$^{-2}$ s$^{-1}$) at the photon-ring scale and detect Sgr A*'s quiescent emission ($\sim 2 \times 10^{-14}$ erg cm$^{-2}$ s$^{-1}$) within a single dynamical time ($r_g/c \sim 20$ s). Effective area requirements range from $\gtrsim 1000$ cm$^2$ at 1 keV, $\gtrsim 100$ cm$^2$ at 6.4 keV.
 
Angular resolutions of $\lesssim 1\,\mu$as at 1 keV, with a goal of $0.1\,\mu$as, would enable photon-ring resolution at Fe-K$\alpha$. As can be seen in Figure \ref{fig:jetfeature_res} and Table \ref{tab:feature_sizes}, this resolution would resolve 
$r_g$ for M87 ($3.8\,\mu$as) and Sgr A* ($5.0\,\mu$as) at $\sim 5$--$10$ resolution elements across the gravitational radius. A field of view of $\sim400~\mu$as at full angular resolution is sufficient to image the launching region ($\lesssim 100\,r_g \approx 380\,\mu$as for M87) in addition to the jet base ($\sim 1000\,r_g$) in a single observation run.

This case requires two primary science-driven sub-bands: (i) $\sim0.5-2$ keV soft band for jet synchrotron continuum and resolved corona-jet boundary imaging; (ii) $\sim6-7$ keV hard band centered on Fe-K$\alpha$ at $6.4$ keV for photon ring and shadow imaging where the line-to-continuum ratio enhances the GR signature. An extension to $\sim0.3-15$ keV would enable the capture of the soft excess and Comptonized continuum. Spectral resolutions of $E/\Delta E \sim 50$ baseline ($\Delta E \sim 130$ eV at 6.4 keV) would be sufficient to isolate the Fe-K$\alpha$ line for photon ring imaging. To resolve the Fe-K$\alpha$ line profile and to disentangle gravitational redshift and Doppler broadening from disk rotation requires a more stringent spectral resolution requirement of $E/\Delta E \gtrsim 1000$ ($\Delta E \sim 6$ eV at 6.4 keV), which would enable a direct probe of spin and inclination at the launching scale. 
Further, this case would require $\sim 1$~ms event timing, sufficient to resolve hot-spot orbital motion at $r_{g}/c$ for Sgr A* ($\sim 20$s) and to phase-resolve QPO timescales in nearby AGN ($\sim 100-1000$s).

 
 


Finally, in addition to the specifications noted above, there are several improvements if a polarimetry instrument could be added to the architecture. In particular, as opposed to simply ``inferring" the magnetic field topology and orientation with X-ray imaging data coupled with correlated IXPE observations, a designated polarimetry instrument would enable the direct measurement of these parameters. While further study is required to assess if this instrument can effectively be added to AEI, we highlight the nominal specs that would be needed here: Minimum detectable polarization (MDP) $\lesssim 1\%$ at $99\%$ confidence in $10^4$s for M87-class targets, required for resolved Stokes Q/U mapping of the launching region and for tests of the photon-ring polarization-angle. Sensitivity for Stokes V at the $\sim0.1\%$ level is required for Faraday rotation detections.

\section{Science Traceability Matrix}\label{sec:STM}
\begin{figure}[t!]
  \centering
  \includegraphics[width=0.99\textwidth]{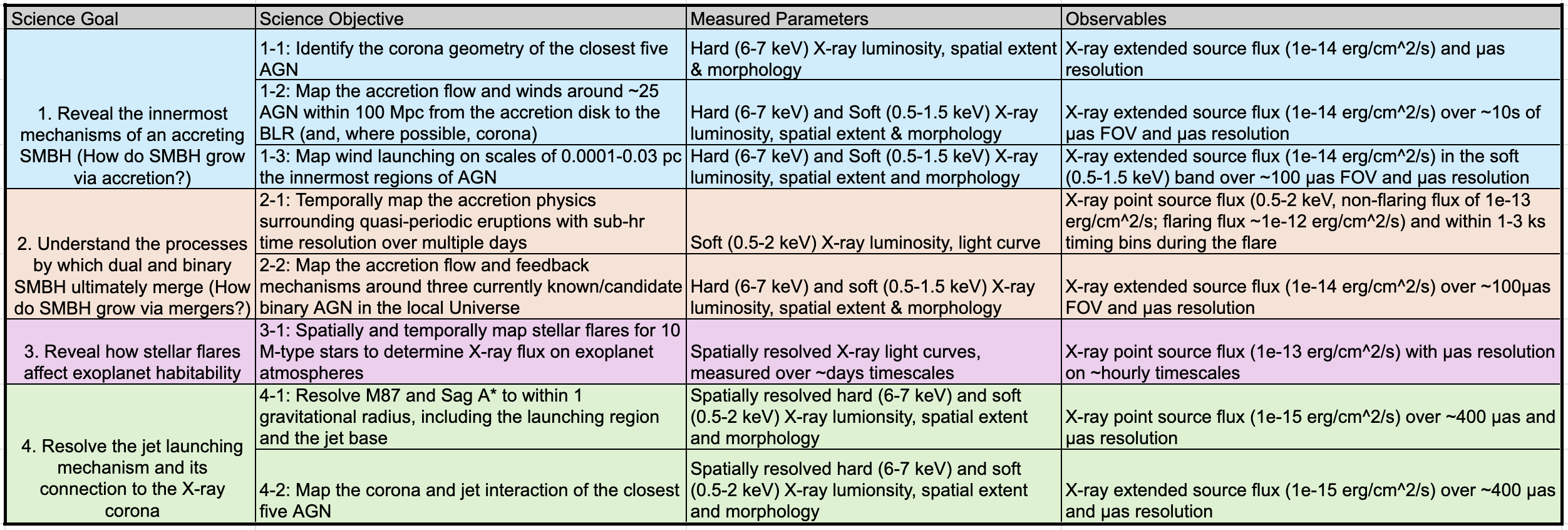} 
    \caption{Science traceability matrix developed for the Accretion Explorer Interferometer mission concept. With an ultra-high resolution X-ray imager such as AEI, we can explore critical open questions regarding the growth and feedback mechanisms of SMBHs, the precursors of gravitational wave events, and the effects of stellar flares on exoplanet habitability.}
     \label{fig:stm}
\end{figure}

\begin{figure}[t!]
  \centering
\includegraphics[width=0.5\textwidth]{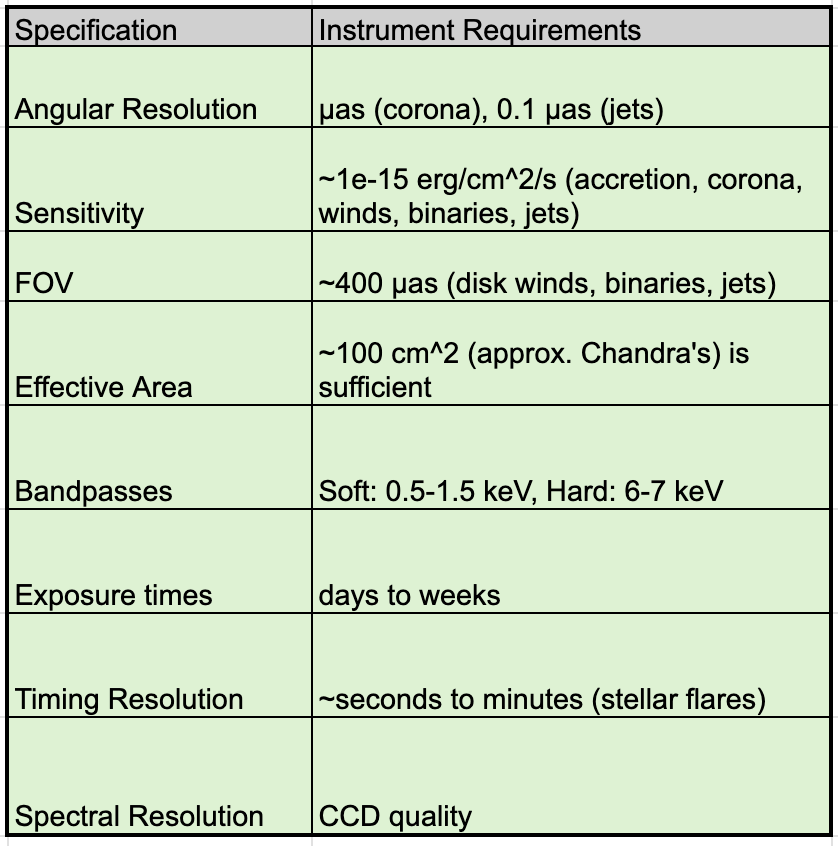}
    \caption{Critical technical specifications to address the science goals delineated in the science traceability matrix. }
     \label{fig:techspec}
\end{figure}

The science traceability matrix for AEI can be found in Figure \ref{fig:stm}. By compiling the most stringent technology requirements to successfully achieve science goals, we identify the ideal instrument requirements in Figure \ref{fig:techspec}. Across all science cases, the highest priority specifications are primarily angular resolution ($0.1-1\mu$as), and sensitivity ($\sim10^{-15}$~erg~cm$^{-2}$~s$^{-1}$ or better; see Section \ref{ssec:questions} for a discussion on uncertainties). While sensitivity estimates were done assuming Chandra's effective area, exposure times for the faintest and extended science targets are on the order of days to weeks, and larger effective area could provide the sensitivity needed for more efficient observing. 

We note that this science traceability matrix takes into consideration only the $\mu$as instrument on board the AEI concept. The two X-ray spotting telescopes are currently planned to have 0.1$''$ and mas resolution, and there is significant discovery space that could be done to investigate multi-scale science questions when taking the full observatory instrument suite into consideration. Multiple concepts for the two X-ray spotting telescopes are currently being explored \citep[e.g.,][]{chalifoux2020}, and the particular instrument specifications are not yet constrained. As these explorations continue, further refinement of the STM to account for the full suite of instruments and demonstrate the full scientific potential of the AEI concept.

\subsection{Open Questions}
\label{ssec:questions}

AEI would provide a transformative look into the X-ray sky. As a result of probing the frontiers of what is possible with this significant technical leap, there are several knowledge gaps that require additional study with theoretical simulations or observations to ensure that our science cases are robustly developed. 

\subsubsection{Science Hurdles}

One notable uncertainty is within the calculation of sensitivity requirements. As we only currently have integrated luminosities over arcsecond-scale resolutions, there is significant uncertainty in the flux distribution when probing smaller scales. This is particularly true within the central engine of the AGN where there are still major outstanding questions regarding the geometry of the accretion disk, X-ray corona, and jet launching region. Therefore, the sensitivities needed to probe certain regions are highly uncertain. Throughout the text, sensitivities were calculated using different mechanisms to define a reasonable range to be explored. With stepwise resolution improvements for X-ray imagers (e.g., a mas-resolution imager) and subsequent validation of models of these small regions, these estimates can be better constrained.


A further source of uncertainty is in the sample for binary supermassive black holes. As discussed earlier, there is currently only one robustly confirmed SMBH binary that has been identified to date \citep{rodriguez2006}. While LISA is predicted to detect dozens to hundreds more in its lifetime, poor sky localization and short lead time to coalescence will prove difficult for AEI to react directly to LISA detections as observations must be months or more in advance within the current proposed architecture and concept of operations. Therefore, an improved understanding of the electromagnetic signatures of binary supermassive black holes, and the compilation of a robust sample of binary AGN is critical to the successful completion of the binary science case. There are several promising identification diagnostics that could provide the needed sample size. In addition to radio VLBI which was used to identify CSO 0402+739 \citep{rodriguez2006}, techniques such as astrometric varstrometry \citep{shen2019}, variability \citep{kelley2019}, and the use of Doppler-shifted spectroscopic signatures \citep[e.g.,][]{dotti2023,mondal2024}. Moreover, several proposed, current, and upcoming facilities have noted identifying pc- and sub-pc scale binary systems as a critical science goal, such as the ngVLA \citep{wilner2024}, the Square Kilometre Array \citep{deane2015}, and AXIS \citep{foord2024}, among others \citep[see][for a review]{derosa2019}. Upon further study with current and upcoming facilities, a robust sample of binary sources could be come available.


Finally, there are currently no general relativistic radiative transfer (GRRT) codes that ray-trace simulated images produced by self-synchrotron or inverse Compton emission/absorption models at high-energy. Additionally, the computational pipeline for simulating accreting Kerr black holes with jet emission models remains disconnected from GRRT codes. Where GRRT is implemented in the post-processing stage of a standard simulated imaging pipeline. A fully self-consistent idealized ray-tracing code would be needed to forecast simulated images at X-ray energies.

\subsubsection{Technical Hurdles}
A larger discussion of technical hurdles for the AEI concept can be found in the NIAC study report \citep{weaver2026b}. Here we note several open questions and uncertainties regarding mission architecture and operations related to the science goals that must be investigated in future studies. One of the largest open questions concerns the sky coverage of the observatory. To avoid stray light from the Sun contaminating observations and endangering the instruments, there is a significant solar exclusion angle of $\pm20^\circ$ from the ecliptic plane.

Moreover, the slew times and resource intensive $\Delta v$ budget place strong constraints on scheduling. The goal lifetime for the concept is five years, however the NIAC study identified rough estimates of $\Delta v$ constraints placing an upper limit to mission lifetime of $\approx 2$ years. Considering reasonable assumptions for advancements in propulsion technology and improved modeling and planning tools, a five-year primary mission could be within reach, however there are significant scheduling considerations. In particular, to optimize resource use, observations must be strictly scheduled, with each six-month block of observations reserved for a single quadrant of the sky. Targets must be scheduled so that the observatory is continually moving in one direction, to avoid unnecessarily large slews and ``dead time" where no observations can take place.

Due to such stringent scheduling requirements, significant lead time is needed for effective operations. As the observatory is currently designed, primary mission targets must be identified and scheduled pre-launch. Additional study is needed to find improved optimizations to slewing and $\Delta v$, to investigate mitigation strategies to minimize the solar exclusion angle, and to optimize science targets given the constraints.






\section{Conclusions}

Our ability to continue growing our understanding of the high energy environments surrounding supermassive black holes and active stars hinges on continued technology development to open new windows into these extreme processes. In particular, the prospect of ultra high resolution imaging ($\sim$~mas$-\mu$as) would open up new avenues for investigation into the smallest, yet critical, scales to answer some of the most pressing open questions in astrophysics.
The Accretion Explorer Interferometer is one concept that is being explored to provide these capabilities. With $\mu$as-scale resolution, we can achieve five to six orders of magnitude greater than current facilities, opening up an entirely new transformative discovery space into the physics of accretion and jet feedback, the precursors and electromagnetic counterparts of gravitational wave events, and the effects of stellar magnetic activity on exoplanet environments and habitability. We present a science traceability matrix that delineates the key science goals of AEI, and note the most critical technology capabilities, in addition to angular resolution, for the successful completion of these goals, particularly high sensitivity and effective area. As continued advancements in multi-wavelength instrumentation continue to unlock access to smaller and smaller spatial scales, it is critical that X-ray instrumentation can keep up with the high-resolution demands required to answer these large open questions.

\appendix    

\section{Example Target Selection}

Here we present a nominal target sample to probe the science cases discussed above. A more condensed target list that was used to assess $\Delta v$ and observation planning in the NIAC study can be found in the NIAC report \citep{weaver2026b}. 

     

\begin{table}[ht!]
\centering
\caption{Priority targets for Science Case 1: Accretion}
\begin{tabular}{ccccc}
\hline
\hline
Target Name            & RA (degrees) & DEC (degrees) & $\log{(M_{\mathrm{BH}}\,/\,\mathrm{M}_{\odot})}$ & Corona Size\,($\mu$as) \\ \hline
NGC 3998               & 179.483889   & 55.453589     & $8.93$                    & 5.46                          \\
4U1344-60              & 206.900028   & $-$60.617743  & $9.09$                    & 2.22                           \\
4C+74.26               & 310.655450   & 75.134012  & $9.83$                    & 1.66                            \\
Mrk 3                  & 93.901482    & 71.037482     & $8.96$                    & 1.53                           \\
\hline
\end{tabular}
\label{tab:corona}
Priority targets from the BAT AGN Spectroscopic Survey Data Release 2 \citep{bassDR2} to resolve and study the central engine of nearby AGNs (corona, accretion disk, disk winds, torus). These targets have been selected as their X-ray corona can be resolved with $\mu$as resolution; a larger sample to resolve accretion disk and disk winds exists.
\end{table}

\begin{table*}[ht!]
\caption{Priority targets for Science Case 2: Binary AGN and QPEs}
\label{tab:grav_candidates}
\centering
\resizebox{\textwidth}{!}{%
\begin{tabular}{lcccccc}
\hline
Target & science case & RA & DEC & $z$ & pair separation (for binaries) \\
& & & & & recurrence time (QPEs) \\
\hline
CSO 0402+739 & confirmed binary & 04:05:49.2630 & +38:03:32.234 & 0.055 & 7.3 pc \\

PG 0851+202 & binary candidate & 08:54:48.8749 & +20:06:30.640 &  0.306 & 0.1 pc \\
4C+22.25  &  binary candidate & 10:00:21.7953 & +22:33:18.805 & 0.419 & 0.1 pc \\
GSN 069 &  QPE & 01:19:08.4305 & $-$34:11:30.980 & 0.018 & $\sim9$~hr\\
RX J1301.9+2747 &  QPE & 13:02:00.1380 & +27:46:57.849 & 0.024 & $\sim3.6-5.6$~hr \\
eRO-QPE1 & QPE & 02:31:47.1623 & $-$10:20:11.017 & 0.051 & $\sim7.6$~hr \\
eRO-QPE2 & QPE & 02:34:48.9700 & $-$44:19:31.650 & 0.018 & $\sim2.4$~hr \\
eRO-QPE3 & QPE & 14:00:53.3263 & $-$28:46:01.165 & 0.024 & $\sim20.4$~hr \\
eRO-QPE4 & QPE & 04:45:33.8108 & $-$10:12:04.741 & 0.044 & $\sim9.8-14.7$~hr\\
AT2019qiz & QPE & 04:46:37.8885 & $-$10:13:34.968 & 0.015 & $\sim48.4$~hr \\
AT2022upj & QPE & 00:23:56.8460 & $-$14:25:23.220 & 0.054 & $\sim0.5-3.5$~days\\
SDSS1335+0728 & QPE & 13:35:19.91 & +07:28:07.4 & 0.024 & $\sim4.5$~days \\

\hline
\end{tabular}}
 \vspace{0.2cm}
Example priority targets for the binary and QPE science cases. For binary targets, the pair separation is reported, while the typical recurrence times are reported for QPE targets.
\normalsize
\end{table*}

\begin{table*}
\centering
\caption{Priority targets for science case 3: Stellar Activity and Habitability}
\resizebox{\textwidth}{!}{%
\begin{tabular}{lcccccccc}
\hline
Star & SpT & Distance (pc) & $\theta_{\rm diam}$ (mas) & $F_{X,q}$ (erg cm$^{-2}$ s$^{-1}$) & $\log(L_X)$ & Planet Status & Strategy (400 $\mu$as FOV) \\
\hline
Alpha Centauri A+B (combined) & G2V+K1V & 1.35 & 8.56, 5.80 & $9.9\times10^{-12}$ & 27.33 & No confirmed planet & mas-scale / Partial-disk \\
Proxima Centauri & M5.5Ve & 1.30 & 1.10 & $8.4\times10^{-12}$ & 27.23 & Known planet(s) & mas-scale / Partial-disk \\
HD 219134 & K3V & 6.53 & 1.07 & $3.49\times10^{-13}$ & 27.25 & Planet host/candidate (TOI) & mas-scale / Partial-disk \\
AD Leo & M3.5V & 4.97 & 0.79 & $2.5\times10^{-11}$ & 28.87 & No confirmed planet & Partial-disk / Active Region \\
AU Mic & M1Ve & 9.72 & 0.67 & $1.8\times10^{-11}$ & 29.31 & Known planet(s) & Partial-disk / Active Region \\
EV Lac & M3.5Ve & 5.05 & 0.65 & $1.0\times10^{-11}$ & 28.48 & No confirmed planet & Partial-disk / Active Region \\
YZ CMi & M4.5Ve & 5.99 & 0.55 & $5.8\times10^{-12}$ & 28.40 & No confirmed planet & Partial-disk / Active Region \\
Wolf 359 & M6Ve & 2.41 & 0.54 & $1.6\times10^{-12}$ & 27.05 & No confirmed planet & Partial-disk / Active Region \\
HD 304636 & M0V & 10.45 & 0.47 & $4.34\times10^{-13}$ & 27.75 & Planet host/candidate (TOI) & Partial-disk / Active Region \\
HD 260655 & M0.0V & 10.01 & 0.44 & $3.75\times10^{-13}$ & 27.65 & Planet host/candidate (TOI) & Partial-disk / Active Region \\
BD-17 588A & M3V & 6.87 & 0.37 & $1.98\times10^{-12}$ & 28.05 & Planet host/candidate (TOI) & Full-disk \\
AP Col & M4.5Ve & 8.66 & 0.31 & $4.8\times10^{-12}$ & 28.63 & No confirmed planet & Full-disk \\
\hline
\end{tabular}}


\textbf{Nearby stellar targets illustrating angular diameter ($\theta_{\rm diam}$) and X-ray brightness, ordered by decreasing $\theta_{\rm diam}$.} The final column categorizes the optimal observation strategy assuming a hard 400 $\mu$as field of view (FOV) limit; targets $> 0.4$ mas require partial-disk pointings to isolate specific active regions. Quiescent fluxes ($F_{X,q}$) are sourced from established X-ray catalogs (SIMBAD, VizieR, literature) or estimated from ROSAT count rates (0.1--2.4 keV) assuming typical coronal spectra. Flare-state fluxes can exceed these quiescent values by 1--3 orders of magnitude.

\label{tab:stellar_candidates}
\end{table*}

\begin{table}[ht]
    \centering
    \caption{Priority targets for jets science case}

    \begin{tabular}{lcccc}
    \hline
    \hline
        Target & RA & Dec & Photon ring scale & Jet base scale \\
        & & & $\mu$as & $\mu$as \\
        \hline
        \hline
        Sag A$^*$ & 17:45:40.0383 & -29:00:28.069 & 25.7 & 49.4 \\
        M87$^{*}$ & 12:30:49.42 & +12:23:28.04 & 20 & 38.5 \\
        Mrk 501 & 16:53:52.21 & +39:45:36.61 & 0.368  & 0.708 \\
         \hline
         \hline
    \end{tabular}
    \label{tab:placeholder}\\
     \vspace{0.2cm}
Example targets were identified based on goals to resolve the jet base and photon ring via the calculations in Figure \ref{fig:jetfeature_res}.
\end{table}


\subsection*{Disclosures}
The authors declare that there are no financial interests, commercial affiliations, or other potential conflicts of interest that could have influenced the objectivity of this research or the writing of this paper.


\subsection* {Code, Data, and Materials Availability} 
Data sharing is not applicable to this article, as no new data were created or analyzed.


\subsection* {Acknowledgments}
This work is supported by NASA under award number 80GSFC24M0006 (JMC, TM, LDV, RG, MM, KM, EL, MS). 
L.D.V. acknowledges support from the Heising-Simons Postdoctoral Launch Program Fellowship. This work was also supported by NIAC Phase I proposal.


\bibliography{report}   
\bibliographystyle{spiejour}   




\listoffigures
\listoftables

\end{spacing}
\end{document}